\documentclass[%
 aip,
 amsmath,amssymb,
 reprint,%
]{revtex4-1}

\renewcommand{\selectlanguage}[1]{}

\usepackage{graphicx}
\usepackage{dcolumn}
\usepackage{bm}

\usepackage[utf8]{inputenc}
\usepackage[T1]{fontenc}
\usepackage{mathptmx}
\usepackage{etoolbox}
\usepackage{float}

\makeatletter
\def\@email#1#2{%
 \endgroup
 \patchcmd{\titleblock@produce}
  {\frontmatter@RRAPformat}
  {\frontmatter@RRAPformat{\produce@RRAP{*#1\href{mailto:#2}{#2}}}\frontmatter@RRAPformat}
  {}{}
}%
\makeatother
\begin{document}

\preprint{AIP/123-QED}

\title{A variability-aware simulation and design workflow for wafer-scale, heterogeneously integrated lithium niobate modulators}
\author{P.Nenezic}
\email{patrick.nenezic@ugent.be}
\affiliation{ 
Department of Information Technology, Photonics Research Group, Ghent University–imec, 9052 Ghent, Belgium.
}%
\affiliation{%
imec, Kapeldreef 75, 3001 Leuven, Belgium. 
}%

\author{E. Vissers}%
\affiliation{ 
Department of Information Technology, Photonics Research Group, Ghent University–imec, 9052 Ghent, Belgium.
}%
\affiliation{%
imec, Kapeldreef 75, 3001 Leuven, Belgium. 
}%
\author{A. Moerman}%
\affiliation{ 
Department of Information Technology, Photonics Research Group, Ghent University–imec, 9052 Ghent, Belgium.
}%
\affiliation{%
imec, Kapeldreef 75, 3001 Leuven, Belgium. 
}%
\affiliation{%
Department of Information Technology, IDLab, Ghent University–imec, 9052 Ghent, Belgium.
}%
\author{L. Bogaert}%
\affiliation{ 
Department of Information Technology, Photonics Research Group, Ghent University–imec, 9052 Ghent, Belgium.
}%
\affiliation{%
imec, Kapeldreef 75, 3001 Leuven, Belgium. 
}%
\author{M. Billet}%
\affiliation{ 
Department of Information Technology, Photonics Research Group, Ghent University–imec, 9052 Ghent, Belgium.
}%
\affiliation{%
imec, Kapeldreef 75, 3001 Leuven, Belgium. 
}%
\author{X. Zheng}%
\affiliation{ 
Department of Information Technology, Photonics Research Group, Ghent University–imec, 9052 Ghent, Belgium.
}%
\affiliation{%
imec, Kapeldreef 75, 3001 Leuven, Belgium. 
}%
\author{T. Vanackere}%
\affiliation{ 
Department of Information Technology, Photonics Research Group, Ghent University–imec, 9052 Ghent, Belgium.
}%
\affiliation{%
imec, Kapeldreef 75, 3001 Leuven, Belgium. 
}%
\author{M. Niels}%
\affiliation{ 
Department of Information Technology, Photonics Research Group, Ghent University–imec, 9052 Ghent, Belgium.
}%
\affiliation{%
imec, Kapeldreef 75, 3001 Leuven, Belgium. 
}%
\author{A. Papadopoulou}%
\affiliation{ 
Department of Information Technology, Photonics Research Group, Ghent University–imec, 9052 Ghent, Belgium.
}%
\affiliation{%
imec, Kapeldreef 75, 3001 Leuven, Belgium. 
}%
\author{S. Uvin}%
\affiliation{ 
Department of Information Technology, Photonics Research Group, Ghent University–imec, 9052 Ghent, Belgium.
}%
\affiliation{%
imec, Kapeldreef 75, 3001 Leuven, Belgium. 
}%
\author{P. De Heyn}%
\affiliation{%
imec, Kapeldreef 75, 3001 Leuven, Belgium. 
}%
\author{S. Saseendran}%
\affiliation{%
imec, Kapeldreef 75, 3001 Leuven, Belgium. 
}%
\author{S. Atzeni}%
\author{B. Kuyken}%
\affiliation{ 
Department of Information Technology, Photonics Research Group, Ghent University–imec, 9052 Ghent, Belgium.
}%
\affiliation{%
imec, Kapeldreef 75, 3001 Leuven, Belgium. 
}%


\begin{abstract}
We present a variability‑aware simulation framework for heterogeneously integrated lithium niobate traveling‑wave modulators. The framework incorporates fabrication‑variation data obtained from our dedicated pilot line and enables efficient optimisation of geometric parameters to ensure stable device performance across wafer‑scale manufacturing. The proposed multi-parameter optimisation method enables the efficient identification of modulator designs that simultaneously achieve target performance metrics ($V_{\pi}$, optical insertion loss, and 3\,dB electro-optic bandwidth) while maintaining robustness against fabrication‑induced variations. Using this methodology on two representative modulator architectures, we theoretically demonstrate that reliable wafer‑scale integration of lithium niobate modulators on silicon photonics via micro‑transfer printing is feasible and can be systematically engineered.
\end{abstract}

\maketitle

\section{\label{sec:Introduction}Introduction}

The rapid, exponential growth of data traffic is driving power consumption in data centres to unsustainable levels. Silicon photonics (SiPho) platforms have enabled significantly higher data rates with greatly reduced energy requirements. Silicon photonic integrated circuits (Si‑PICs) make it possible to densely integrate complex optical functions on a single silicon chip using standard CMOS fabrication processes, supporting high‑volume, high‑yield, and cost‑effective manufacturing. However, emerging communication standards are now targeting baud rates beyond 200~GBd, levels that are difficult to attain with current silicon photonic modulators. 

To overcome these limitations, several alternative electro‑optic (EO) materials have been investigated, including BTO~\cite{chelladurai_barium_2025}, organic materials~\cite{wolf_silicon-organic_2018}, plasmonics~\cite{haffner_all-plasmonic_2015}, and more recently lithium tantalate~\cite{wang_lithium_2024, niels_high-speed_2026}. Among these, thin‑film lithium niobate (TFLN) stands out due to its relative maturity, its low loss and strong Pockels effect, with monolithic devices already exceeding 100~GHz bandwidth~\cite{zhu_integrated_2021}. However, pure TFLN platforms lack integrated photodetectors and CMOS‑compatible functionality. Heterogeneous TFLN-on-silicon integration addresses this gap by combining EO modulation performance of TFLN with the mature, high‑yield components of silicon photonics and CMOS electronics, enabling a more complete platform.

As lithium‑niobate (LN) modulator technologies continue to mature, an important question is whether this material platform, particularly in its heterogeneously integrated form, can be scaled to support wafer‑level device manufacturing. Several efforts toward such scalability have emerged, primarily relying on wafer bonding~\cite{rahman_integration_2025,churaev_heterogeneously_2023, ghosh_wafer-scale_2023} or micro‑transfer printing (MTP)~\cite{niels_wafer-scale_2025_OL, niels_advances_2026,lu_broadband_2026}. While these studies demonstrate the feasibility of wafer‑scale processing, they do not address the design constraints needed to ensure that modulators maintain their target performance specifications under high‑yield wafer‑scale fabrication. 
Achieving robust, reproducible device performance at wafer scale demands modulator designs that can tolerate fabrication‑induced variations. 
Although variability‑aware design methodologies have been explored for silicon photonic circuit components~\cite{bogaerts_layout-aware_2019, zhang_decomposed_2020, ullrick_wideband_2023}, they have not been extended to traveling-wave electro‑optic modulators, and in particular not to heterogeneously integrated LN modulators. 

First, this work proposes two modulation architectures compatible with wafer-scale MTP and provides a comparative analysis. Furthermore, it establishes a unified EO design framework that allows for the efficient optimisation of heterogeneously integrated LN modulators while explicitly incorporating fabrication-induced variability into the design process. By using fabrication variation data from our TRANSVERSE pilot line, we demonstrate that, with a rigorous and systematic design methodology, the wafer‑scale integration of LN modulators on silicon photonics using micro‑transfer printing is feasible. 

In Section 2, based on our recent results of micro-transfer printed 200~mm silicon photonics wafer, we propose two design architectures which act as the example of our analysis. Section 3, with the help of a created simulation database, analyses the influence of geometric parameters on device performance. Finally, Section 4 introduces a new design framework and presents a statistical optimisation approach that incorporates fabrication variations into the optimisation process. Although the proposed framework is demonstrated with micro-transfer printing, it is equally applicable to other heterogeneous integration approaches such as die-to-wafer bonding. Furthermore, due to similarities in material properties, the presented methodology is also valid for heterogeneous lithium tantalate traveling wave modulators.

\begin{figure*}
\includegraphics[width=0.9\textwidth]{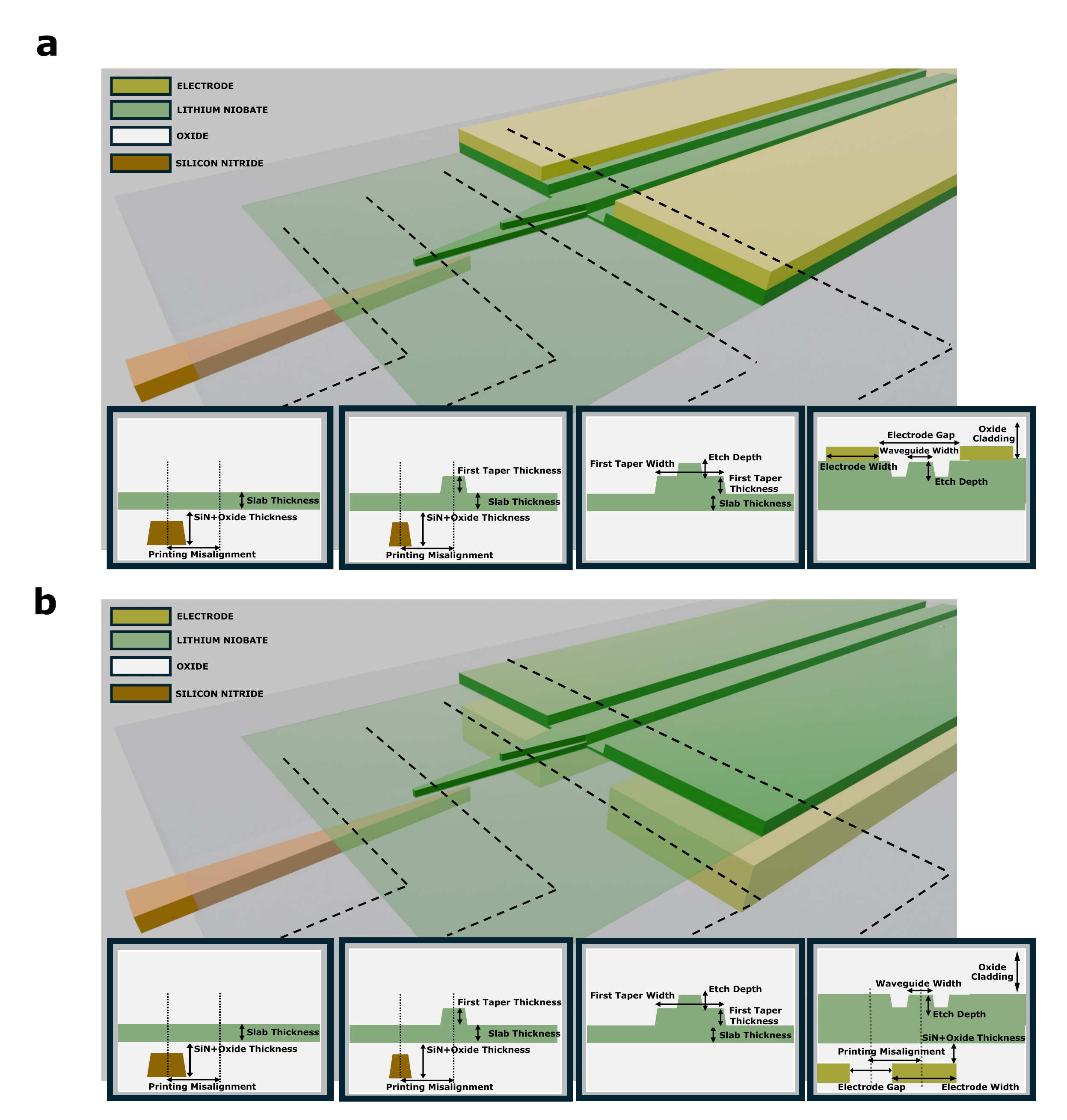}
\caption{\label{fig:bottomandtop} (a) Schematic of the top electrode design. Metal electrodes are fabricated on the lithium niobate prior to integration onto the silicon nitride target wafer. This approach prevents the accumulation of alignment errors between the lithium niobate placement and the electrode definition, which can occur in post-processing metallisation schemes. Light is coupled from the target wafer into the lithium niobate waveguide via a two-layer adiabatic taper, minimising optical losses. (b) Schematic of the bottom electrode design. Metal electrodes are embedded in the target wafer, allowing an unmetallised lithium niobate layer to be integrated. This eliminates the need for post-processing metallisation and, similarly to the top electrode configuration, avoids cumulative alignment errors. The same tapering approach is applied as in the top electrode design.}
\end{figure*}

\section{\label{sec:CurrentState}Current state of micro transfer printing at wafer scale and proposed device stacks}

Recently, we demonstrated reliable MTP of LN waveguide-based modulators onto a full 200~mm silicon photonics wafer~\cite{zheng_micro-transfer_2026}. However, performance limitations arise from the use of hybrid SiN/LN optical modes. To achieve improved performance in next-generation modulators, we propose two device architectures, as show in Fig.~\ref{fig:bottomandtop}. 

First, to enhance the EO efficiency of heterogeneously integrated LN modulators, we replace hybrid confinement in SiN/LN structure with full mode confinement in LN.
This would allow the optical mode to have a greater fraction inside the EO material than in hybrid architectures, thus experiencing a stronger electro-optic interaction. In order to achieve full coupling into LN waveguide, light is coupled from the silicon nitride, located in the target wafer, into the micro-transfer printed lithium niobate waveguide via multi-layer adiabatic waveguide tapering as shown in both  Fig.~\ref{fig:bottomandtop}(a) and Fig.~\ref{fig:bottomandtop}(b). In Fig.~\ref{fig:bottomandtop}(a), i.e. top electrode configuration, metallisation of the modulator is performed on the lithium niobate before integration. Contrary to the post-metallisation schemes, the metal is not defined with respect to the target wafer but with respect to the LN waveguide. This ensures printing misalignment of the lithium niobate is not added to the metal misalignment as it would with post-printing metallisation. This represents a second key difference compared to the method reported by Zheng and coworkers~\cite{zheng_micro-transfer_2026}. It is worth noting that a post-metallisation step may still be required for metal routing and bondpads defintion. However, this step imposes less stringent alignment requirements, as the metal is located far from the modulation region and optical transitions. In Fig.~\ref{fig:bottomandtop}(b) the modulating section is different: the electrode is buried into the target wafer, meaning that once the lithium niobate structure is printed, no metallisation is required. This configuration, i.e. bottom electrode, has not yet been widely adopted, although it has been proposed in literature as an alternative to the more standard approaches with post-processing metallisation~\cite{safian_foundry-compatible_2020, boynton_heterogeneously_2020,zhang_integrated_2021}. A great advantage of such configuration is that the metallisation can take place in a CMOS environment, consequently reducing overhead of back-end processing of heterogeneous integration, let it be micro-transfer printing or wafer bonding. Furthermore, it eliminates cumulative alignment errors due to LN integration and electrode lithography.

To evaluate the performance of the two proposed architectures, an extensive device performance database was generated by simulating tens of thousands of parameter combinations. Various software tools were then used to extract the corresponding performance metrics, including $V_{\pi}L$, optical propagation loss in the modulating section, tapering loss, and several RF parameters, enabling the determination of the 3 dB bandwidth for each simulated device. The investigated parameters are denoted in the cross sections of Fig.~\ref{fig:bottomandtop}. In this paper, we limit the analysis to an optical wavelength of 1310~nm and assume copper as the metal layer, with material properties fixed to intrinsic ones of bulk copper. In addition, a GSGSG (ground-signal-ground-signal-ground) electrode structure on X-cut LN is adopted for both architectures. It is important to note that these assumptions do not compromise the generality of the proposed method, as variations in these parameters would simply extend the explored parameter space. In particular, we considered only parameters that can be directly tuned within our process flow or within the fabrication processes of our partners. Therefore, the ranges of the geometrical parameters are defined within realistic, fabricable limits. It follows that the set of simulated geometry combinations is closely grounded in realistic design rules and includes structures that are compatible with fabrication in our dedicated pilot line. Additional information on figure of merits, theory, and simulation approaches used for data collection can be found in the Supplementary Material.  

\section{\label{sec:Correlation}Global correlation and Sobol Variance Analysis}

\begin{figure*}
\includegraphics[width=\textwidth]{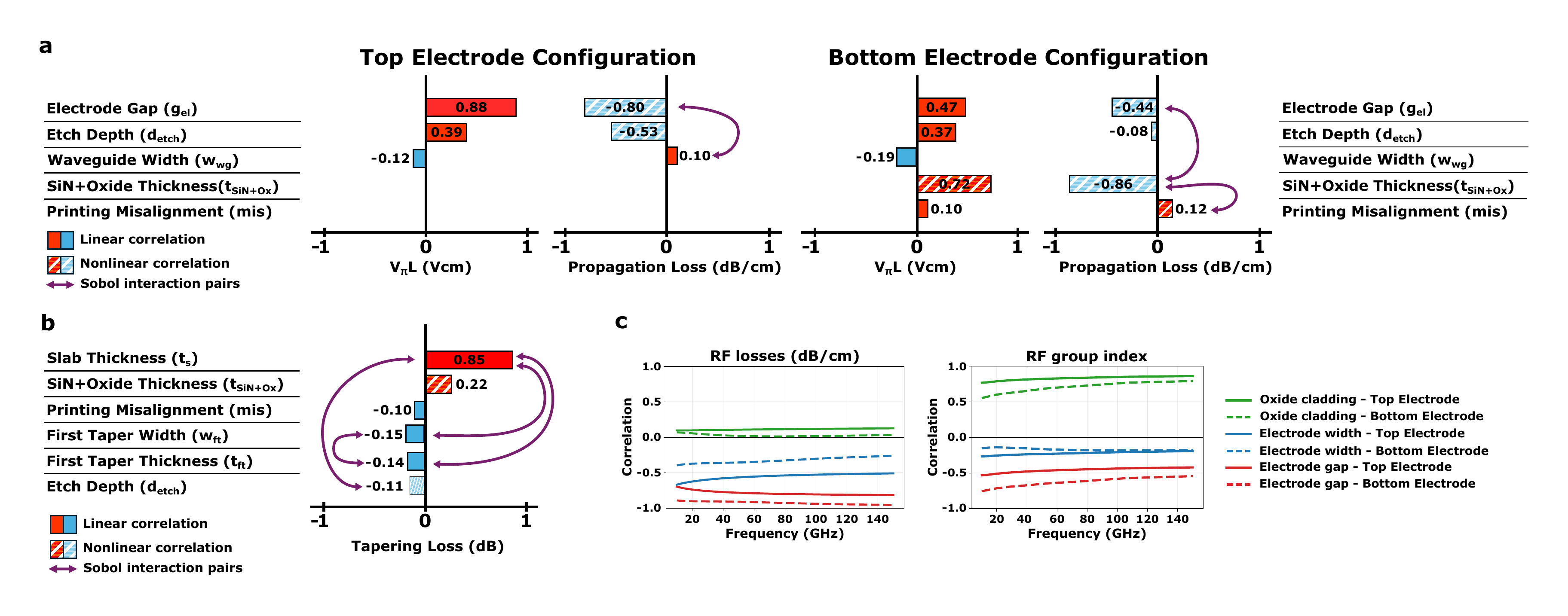}
\caption{\label{fig:correlation}(a) Influence of geometrical parameters on $V_{\pi}L$ and optical propagation loss for top and bottom electrode configurations. The bar plot indicates linear (solid) and nonlinear (striped) correlations, while the purple arrows denote pairwise interactions. (b) Influence of geometrical parameters on tapering loss, with equal results for both top and bottom electrode configurations. (d) Influence of geometrical parameters on RF losses and RF group index for top (solid line) and bottom (dashed line) electrode configurations.}
\end{figure*}

Once the device performance database has been established, it is possible to identify which geometric parameters govern device performance across the whole design
space. To this end, we employ statistical analysis techniques, namely correlation analysis and Sobol variance decomposition. Correlation measures the direction of the output performance metric with respect to a single geometric parameter change, i.e. how changing a geometry parameter improves or worsens the performance metric~\cite{lee_rodgers_thirteen_1988}. On the other hand Sobol indices quantify how much of the total variance of the performance metric is explained by individual parameters (first-order $S_1$) and by their pairwise interactions (second-order $S_2$)~\cite{hart_efficient_2017}. As correlation analyses cannot capture interactions between parameters and Sobol indices do not convey information on trend direction, the two approaches provide complementary insights.
The resulting trends are summarized in Fig.~\ref{fig:correlation}, which provides a compact overview of both dominant parameters and key interactions. Details can
be found in the Supplementary Material.

\subsection{\label{sec:level2}Top electrode configuration}
This paragraph highlights the relationship between device performance and geometrical parameters for top electrode configuration. Figure~\ref{fig:correlation}(a) shows that the electrode gap ($g_{el}$) is the dominant parameter governing the electro-optic figure of merit $V_{\pi}L$, exhibiting a strong linear correlation. In contrast, optical propagation loss follows an opposite trend, decreasing with increasing $g_{el}$ and etch depth ($d_{etch}$), reflecting reduced metal-induced absorption and reduced optical field overlap with the electrode. Printing misalignment has no influence on both $V_{\pi}L$ and propagation loss, as the optical mode is fully confined in the LN waveguide.

While several parameters exhibit largely independent contributions, Fig.~\ref{fig:correlation} also reveals the presence of nonlinear interaction effects. In particular, interactions between waveguide width ($w_{wg}$) and $g_{el}$ indicate that the impact of one parameter can depend strongly on the value of another, leading to non-additive behaviour. For example, when the $g_{el}$ is small, even a modest change in $w_{wg}$ can produce a disproportionately large change in propagation loss, owing to the increased sensitivity of the optical mode to lateral confinement near the metal boundary. The same magnitude of width variation leads to a much weaker loss change when the electrode gap is large. These interaction effects contribute significantly to propagation loss variability and highlight the limitations of treating parameters independently. Such non-additive interactions are denoted in Fig.~\ref{fig:correlation} as Sobol interaction pair arrows.

In addition to metal-induced optical losses, device performance in terms of optical losses depends on how efficiently light is coupled from the SiN waveguide to the LN waveguide. This contribution is referred to as tapering loss. As shown in Fig.~\ref{fig:correlation}(c), tapering loss is primarily governed by the slab thickness $t_s$, with additional weaker dependencies on taper geometry. Strong interaction effects are observed between $t_s$ and the first taper thickness $t_{ft}$, as well as between $t_s$ and $d_{etch}$ indicating that taper optimisation cannot be treated independently. 

The RF properties shown in Fig.~\ref{fig:correlation}(d) are mainly determined by $g_{el}$, electrode width, and oxide cladding thickness. Larger gap and electrode width reduce RF losses, while increased oxide thickness raises the RF group index, indicating slower RF mode propagation. In this configuration, RF-related parameters act largely independently, with limited interaction effects.

\subsection{\label{sec:level2}Bottom electrode configuration}

In the bottom electrode configuration, the correlation dynamics of \(V_{\pi}L\) and propagation losses differ significantly from those of the top electrode configuration, as shown in Fig.~\ref{fig:correlation}(a). For $V_{\pi}L$, both the $g_{el}$, $d_{etch}$ and $t_{SiN+Ox}$ are all strongly correlated parameters, with the latter exhibiting the strongest influence. Increasing any of these geometrical parameters results in a higher \(V_{\pi}L\), and thus a deterioration in device performance. Furthermore, $t_{SiN+Ox}$ correlates with $V_{\pi}L$ in a nonlinear fashion. This is due to the fact that the density of electric field lines overlapping with LN drops nonlinearly if the vertical distance between the electrode and LN increases. Although more parameters influence \(V_{\pi}L\), they affect it independently, similarly to the top electrode configuration.

The correlation analysis of propagation loss reveals that $t_{SiN+Ox}$, which has no influence in the top electrode configuration, exhibits the strongest (nonlinear) correlation in the bottom electrode configuration. Its impact is nearly twice that of $g_{el}$, highlighting this parameter as the dominant factor governing propagation loss. Furthermore, Sobol analysis shows a strict interaction between the two, meaning that they reinforce each others effect. Since the buried electrodes are laterally farther from the waveguide, the EO overlap is less affected by $d_{etch}$ than in the top electrode architecture.

Tapering losses remain unchanged compared to the top electrode configuration, as they are determined prior to the modulation section. RF trends follow similar
behaviour as in the top electrode case, with quantitative differences arising from the modified electrode placement in the device stack.

\subsection{\label{sec:level2}Summary}

The performed analysis shows that device performance is governed by the simultaneous presence of linear, nonlinear, and interaction effects among multiple geometric parameters, thereby rendering the complete optimisation of device performance a highly non-trivial and complex task. Furthermore, when the modulator length is included as an additional degree of freedom, the complexity further increases and additional trade-offs emerge, as longer devices reduce $V_{\pi}$ but increase optical and RF losses and complicate velocity matching.

\begin{figure*}
\includegraphics[width=1\textwidth]{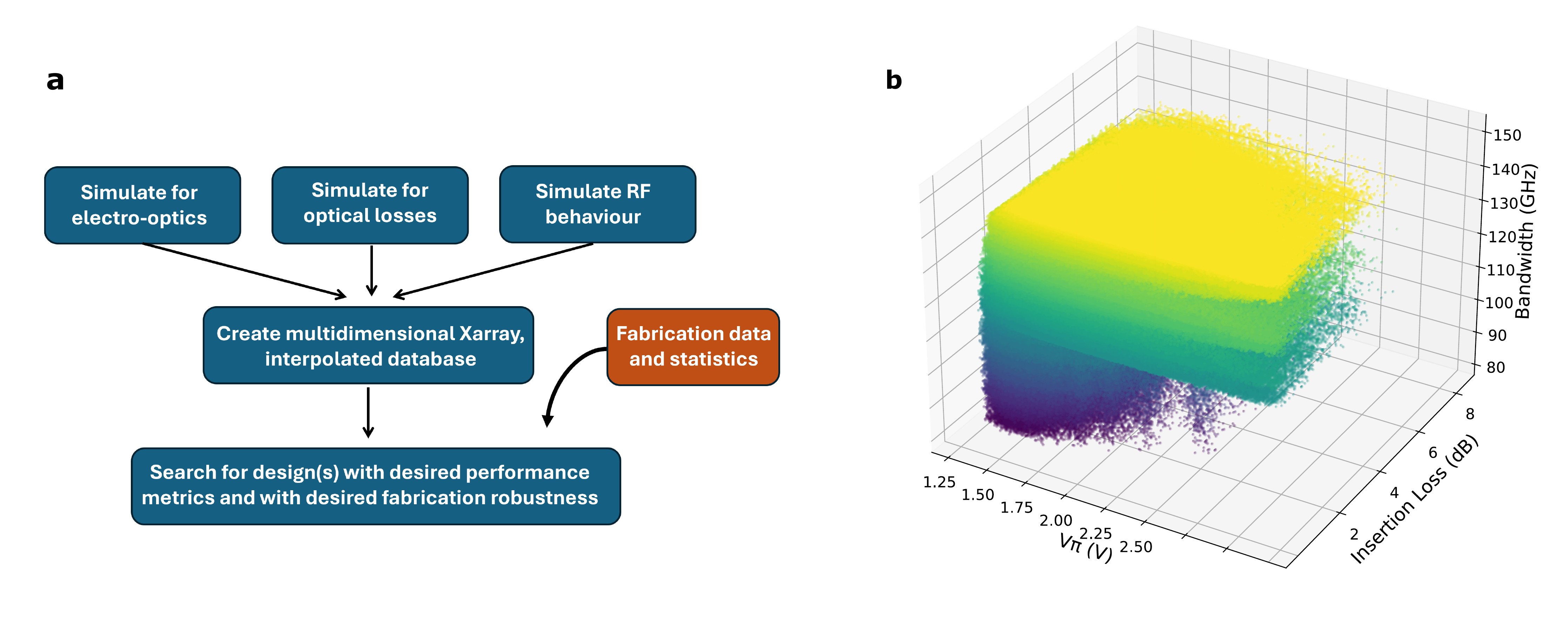}
\caption{\label{fig:cloud} (a) Schematic of the optimisation workflow of the proposed design approach. (b) Three-dimensional projection of the device performance space derived from the XArray database for a 7~mm-long bottom-electrode configuration with GSGSG electrode placement. By varying the modulator length, the distribution of designs shifts and varies within this 3D space. Each point represents a distinct modulator design, and the optimisation is performed within this design space.}
\end{figure*}

These trade-offs and the complexity of geometric parameter effects, further aggravated by pairwise parameter interactions, shows the individual optimisation of parameters is far from optimal, particularly in the bottom electrode configuration. This leads to the need to co-optimise parameters such that \(V_{\pi}\), loss, and bandwidth of a modulator are optimised simultaneously, without requiring iterative optimisation of the device for individual performance metrics. In case fabrication variations are also introduced, the need for a unified and comprehensive device optimisation framework becomes even more pronounced. 
This motivates the optimisation framework introduced in this paper, which identifies optimal modulator designs that simultaneously satisfy all desired performance metrics while accounting for fabrication variations. The optimisation workflow is illustrated in Fig.~\ref{fig:cloud}(a). The first row of the schematic presents the database generation process described in the previous paragraph, while the subsequent steps are detailed in the following sections.


\section{\label{sec:Optimisation}Unified, fabrication aware optimisation method}

The results of the multidimensional simulation sweeps are stored using the Xarray library in Python. Xarray is an open‑source Python package that provides data structures and analytical tools for working with multidimensional labeled arrays, enabling efficient storage, slicing, and interpolation of large scientific datasets~\cite{hoyer_xarray_2017}. Because the database grids are sufficiently dense in
all design dimensions, performance metrics can be evaluated continuously across the full parameter space via interpolation. For any
given device stack, the interpolated dataset can be visualized as a three-dimensional projection in performance space, where the \(x\), \(y\), and \(z\) coordinates correspond to the half-wave voltage $V_{\pi}$, the optical insertion loss, and the RF bandwidth. Such a projected 'cloud' for the bottom electrode configuration is shown in Fig.~\ref{fig:cloud}(b). Here, given a modulator length \(L\), $V_{\pi}$ refers to the
half-wave voltage in a push-pull amplitude modulation configuration, the total insertion loss is given by the propagation loss times \(L\) plus twice the tapering loss, and the bandwidth corresponds to the electro-optic-electric (EOE) 3dB bandwidth. Since these metrics depend on the modulator length, varying the length results in a change and translation of the cloud in the performance space.

Once the database has been generated and interpolated, and the target device performance has been defined, the method can be used not only to identify geometry parameters that achieve the desired performance, but also to determine parameter sets that are robust against unavoidable variations arising from standard fabrication tolerances. For example, in our micro-transfer printed wafer‑scale demonstration, we observed noticeable device‑to‑device performance variations across the wafer~\cite{zheng_micro-transfer_2026}. Therefore, accounting for process variations in the design optimisation routine is crucial when the target performance is required not only at the die level (i.e., for a “hero device”), but also across wafer-scale manufacturing.

Because the database is interpolated and effectively continuous, it is well suited for Monte Carlo analysis. Random perturbations can be drawn from experimentally measured fabrication distributions, and each sample can be evaluated by direct lookup rather than by performing new simulations. This approach significantly reduces the computational cost: a Monte Carlo analysis with 10\,000 samples can be completed in only a few minutes.


\begin{figure*}
\includegraphics[width=1\textwidth]{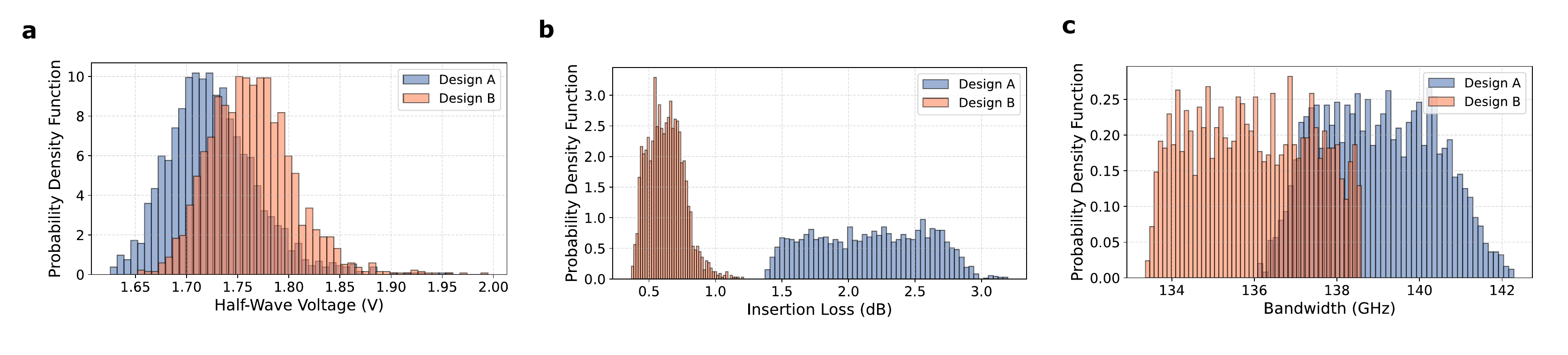}
\caption{\label{fig:comparisonexample} Performance metric comparison of two bottom-electrode designs, presented as probability density functions of the corresponding performance metrics. (a) \(V_{\pi}\) for Designs A and B: both exhibit very similar nominal values and comparable sensitivity to fabrication variations. (b) Insertion loss: Design~B shows lower losses and improved robustness to fabrication variations.  (c) Bandwidth: both designs have similar sensitivity to fabrication variations, with Design~A exhibiting a slightly higher nominal bandwidth.}
\end{figure*}

To highlight the importance of co-optimising performance metrics under fabrication tolerance conditions, two bottom electrode, 7~mm-long modulator designs, denoted as A and B, are evaluated under identical fabrication variation ranges. The results of the Monte Carlo analysis are reported in Fig.~\ref{fig:comparisonexample}. While the same evaluation can also be applied to top electrode configuration, the bottom electrode architecture is selected here as a representative example. Taking into account only the nominal \(V_{\pi}\) values, 1.73~V for A and 1.79~V for B, it would appear that the two designs exhibit nearly identical performance. Furthermore, both designs demonstrate robust behaviour under fabrication variations, as shown in Fig.~\ref{fig:comparisonexample}(a). However, when insertion loss is considered simultaneously, Design~B offers a better overall performance trade-off: it exhibits much lower insertion loss with reduced variability, as shown in Fig.~\ref{fig:comparisonexample}(b). As reported in Fig.~\ref{fig:comparisonexample}(c), Design~A exhibits a slightly larger bandwidth, with a mean value of around 139~GHz in contrast to 136~GHz for Design~B. For most applications, this generally does not justify a penalty of about 1.5~dB in insertion loss. This example highlights the importance of evaluating fabrication variations as well as all performance metrics together in a multidimensional framework rather than relying on any single metric. These observations underscore the need for an automated method to identify geometries that optimally satisfy the target values for \(V_{\pi}\), insertion loss, and bandwidth, while simultaneously maximising robustness against fabrication variations. The method is described in detail in the following paragraph.



Let $\mathbf{x}$ denote the vector containing the nominal geometric parameters of the desired modulator design. However, in practice, the fabricated device may deviate from these nominal values. Therefore, the geometry vector including fabrication variations can be written as

\[
\tilde{\mathbf{X}} =
\begin{pmatrix}
p_1 + \Delta_{1} \\
p_2 + \Delta_{2} \\
p_3 + \Delta_{3} \\
\vdots \\
p_P + \Delta_{P}
\end{pmatrix},
\]


where $\Delta_i$ represents the fabrication-induced deviation of parameter $p_i$ and $P$ is the total number of parameters.
Each deviation has a variation range and follows a probability distribution. These have been obtained from our value chain partners and our own TRANSVERSE MTP pilot line. For example, the MTP process exhibits a quasi-Gaussian misalignment distribution with a $3\sigma$ of $\pm 500\,$nm. This experimentally obtained distributions are embedded directly into the model, with only the nominal vector \(\mathbf{x}\) being adjusted during optimisation, whereas the distributions of $\Delta_i$ remain fixed.

With the probability distributions and ranges defined, a Monte Carlo analysis with $N$ samples can be performed. Then, for a given modulator length, the empirical expected value of each metric \( m \) is computed as

\begin{equation}\label{expectedvalue}
    \hat{\mathbb{E}}[m] = \frac{1}{N} \sum_{k=1}^{N} m\!\left(\tilde{\mathbf{X}}^{(k)}\right),
    \quad m \in \{V_{\pi}, \mathrm{Loss}, \mathrm{BW_{3dB}}\}.
\end{equation}

The expected value represents the average performance of the modulator under fabrication variations. It therefore reflects the typical (mean) metric value one can expect from real fabricated devices of the chosen geometry.

We estimate the variance of a metric $m(\tilde{\mathbf{X}})$ under fabrication variations by

\begin{equation}\label{variance}
\widehat{\operatorname{Var}}[m]
=
\frac{1}{N-1}
\sum_{k=1}^{N}
\big(m(\tilde{\mathbf{X}})-\hat{\mathbb{E}}[m]\big)^2.
\end{equation}
The squared deviations quantify how strongly $m$ fluctuates around its mean, with large outliers penalised more heavily and such that contributions from positive and negative deviations do not cancel each other out. The division by $N\!-\!1$ (Bessel’s correction) accounts for the fact that the sample mean $\hat{\mathbb{E}}[m]$ is estimated from the same data, using up one degree of freedom; this yields an unbiased estimate of the true variance and prevents systematic underestimation for finite $N$. 
Overall, a low variance indicates that the metric is stable, i.e., robust to parameter deviations, whereas a high variance signals that small fabrication errors can cause large performance fluctuations. Thus, variance directly measures fabrication sensitivity. It is worth noting that the relative effect of geometric parameters on the device may differ from the relative correlations observed in Section~2. In that case, the full simulation space was analysed, highlighting global trends, whereas here the fabrication variation ranges and their associated probability distributions determine the actual magnitude of their impact on the performance metrics.

To jointly meet desired performance targets while ensuring robustness to fabrication variations, the following function \(J(\mathbf{x})\) is defined:

\begin{widetext}
\begin{equation}
J(\mathbf{x}) =
\sum_{m \in \{V_{\pi},\, \mathrm{Loss},\, \mathrm{BW}_{3\mathrm{dB}}\}}
\left[
    \alpha_m
    \left(
        \frac{
            \widehat{\mathbb{E}}\!\left[m(\tilde{\mathbf{X}})\right] 
            - m_{\mathrm{target}}
        }{
            R_m
        }
    \right)^2
    \;+\;
    \gamma_m
    \left(
        \frac{
            \widehat{\mathrm{Var}}\!\left[m(\tilde{\mathbf{X}})\right]
        }{
            R_m^{2}
        }
    \right)
\right].
\label{eq:costfunction}
\end{equation}
\end{widetext}


\begin{figure*}
\includegraphics[width=1\textwidth]{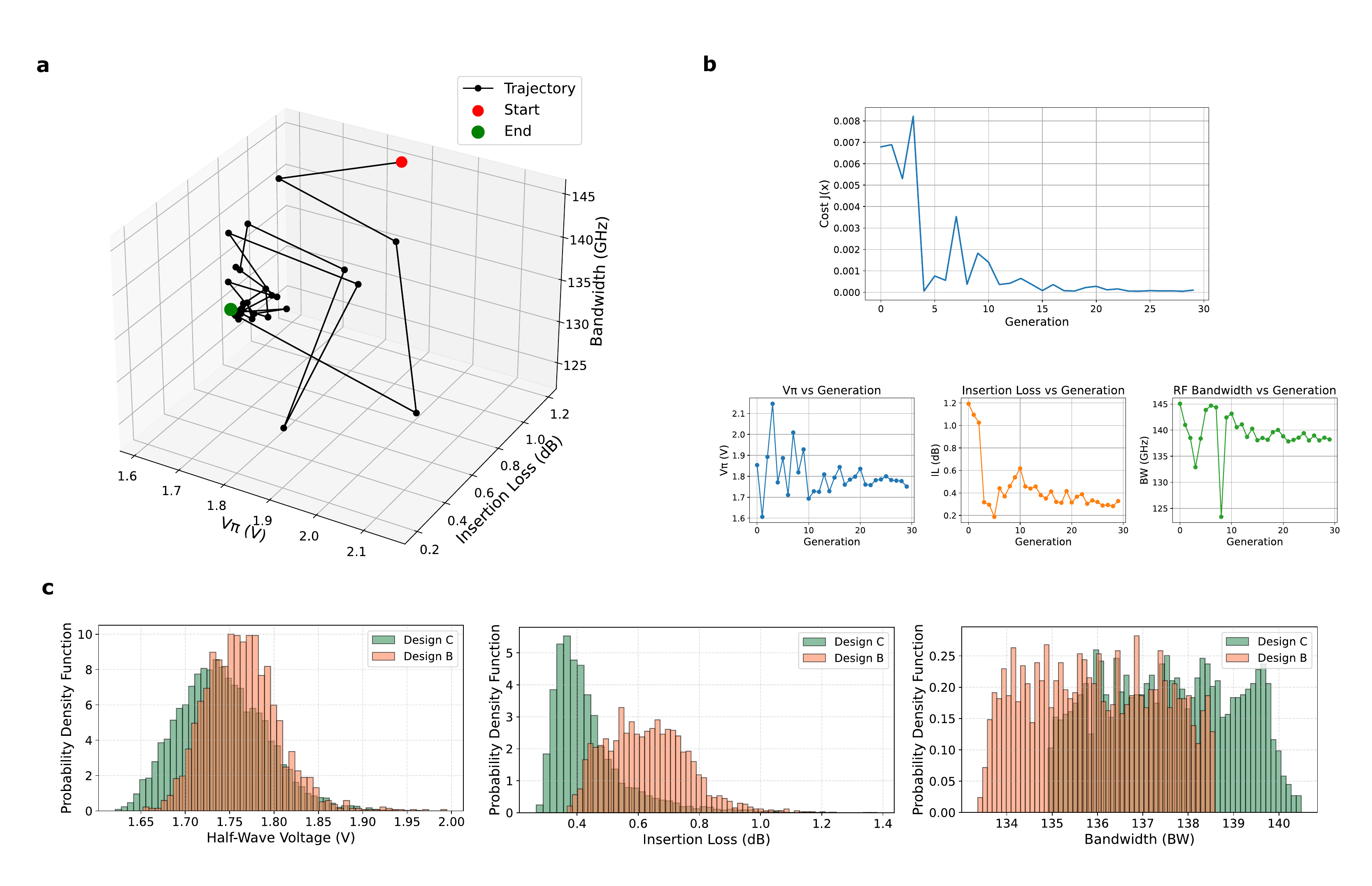}
\caption{\label{fig:cmaes} (a) CMA-ES optimisation trajectory in the performance space: each point represents the best solution at a given generation of statistical simulations, showing the progressive exploration and convergence of the search distribution towards the target metrics. (b) Evolution of the cost function \(J(\mathbf{x})\) and the target metrics \(V_{\pi}\), insertion loss (IL), and 3~dB bandwidth (BW) as a function of the generation. (c) Optimised design comparison: by setting target metrics of Design~B with a priority on robustness, the algorithm identifies Design~C, which achieves improved nominal performance while maintaining comparable or enhanced robustness against fabrication variations.}
\end{figure*}

Here, $m_{\mathrm{target}}$ denotes the desired value of metric $m$, and $R_m$ is the  total range of the given metric within the database. Normalizing the mean–deviation term by $R_m$ and the variance term by $R_m^{2}$ ensures that all three metrics contribute to the objective in a dimensionless and comparable manner, regardless of their physical units or numerical ranges.


\begin{table*}[t!]
\caption{\label{tab:designs} Performance metrics of selected modulators with both bottom and top electrode configurations, designed using the proposed methodology. The reported ranges correspond to the 99th percentile, indicating that 99\% of fabricated devices are expected to fall within these intervals under production conditions. Subscripts BE and TE in column~1 refer to bottom and top electrode configurations, respectively.}
\begin{ruledtabular}
\begin{tabular}{cccccc}
Case &Modulator Length (mm) & V$_\pi$ (V) & Insertion loss (dB) & Bandwidth (~GHz) & Notes \\
\hline
$1_{BE}$ (Design C)&7 & 1.68-1.82 & 0.38-0.62   & 135.5-139.8 &  BiCMOS driving \\
$1_{TE}$&7 & 1.72-1.85 & 1.0-1.55 & 150+ & BiCMOS driving \\
$2_{BE}$&5 & 2.19-2.37  & 0.39-0.64  & 150+  & BiCMOS driving, shorter device \\
$2_{TE}$&5 & 2.2-2.6 & 0.73-0.9 & 150+ & BiCMOS driving, shorter device \\
$3_{BE}$&12 & 0.91-1.0 & 0.48-1.05 & 92-98 & CMOS driving \\
$3_{TE}$&12 & 0.9-1.22 & 2.10-2.48 & 113-116.4 & CMOS driving \\
\end{tabular}
\end{ruledtabular}
\end{table*}

The coefficients $\alpha_m$ and $\gamma_m$ control the relative importance assigned to performance accuracy and fabrication robustness,
respectively. To separate these two roles cleanly, we introduce priority weights $w_m$ that express the designer-selected
importance of matching the target value for metric $m$, and robustness weights $c_m$ that specify how strongly deviations due
to fabrication variations should be penalised. These weights define the cost coefficients as
\begin{equation}
    \alpha_m = w_m,
    \qquad
    \gamma_m = c_m\,\alpha_m .
\end{equation}
%


This formulation ensures that, when $w_m$ are chosen equal, all three target metrics are equally important to reach regardless of the difference in the physical range of the metrics. At the same time, the parameters $c_m$ provide the flexibility to penalise fabrication sensitivity differently across the metrics (e.g.\ enforcing stricter robustness for insertion loss than for bandwidth). For a pre-set modulator length, and given the inputs $m_{\mathrm{target}}$, $\alpha_m$ and $\gamma_m$, the optimal device geometry can be found by minimising $J(\mathbf{x})$. This optimisation is performed using the Covariance Matrix Adaptation Evolution Strategy (CMA-ES). CMA-ES is a gradient-free optimisation method that works by sampling many candidate designs, selecting the best-performing ones, and updating its search distribution to move towards more favourable regions of the design space \cite{nomura_cmaes_2026}. It adapts both the mean and the covariance of its search distribution, allowing it to efficiently explore complex and noisy objective landscapes such as the one analysed. We report an optimisation run in which the performance targets are set to the mean values of Design~B in Fig.~\ref{fig:comparisonexample}. The optimisation trajectory of CMA-ES is shown in Fig.~\ref{fig:cmaes}(a), while Fig.~\ref{fig:cmaes}(b) reports the evolution of the cost function \(J(\mathbf{x})\), as well as the target metrics \(V_{\pi}\), insertion loss, and 3~dB bandwidth, as a function of the generation. The performance metrics of the fully optimised geometry, Design~C, are shown and compared to Design~B in Fig.~\ref{fig:cmaes}(c). Design~C outperforms the initial design in terms of all nominal metrics, while maintaining comparable or improved robustness to fabrication tolerances.


To demonstrate the effectiveness of the methodology, we investigated several configurations compatible with the processing workflow established in the TRANSVERSE pilot line. In particular, the optimisation routine was applied to identify the parameters of heterogeneously integrated LN modulators that meet predefined performance targets, as detailed below. Case~1 is based on a standard modulator length, as reported in~\cite{zheng_micro-transfer_2026}, and aims to achieve low and robust insertion loss, a half-wave voltage compatible with BiCMOS driving, and a high bandwidth exceeding 110~GHz. Case~2 targets a shorter device reaching an even higher bandwidth above 150~GHz, while minimising insertion loss and maintaining \(V_{\pi}\) compatibility with BiCMOS driving. Case~3 focuses on achieving a \(V_{\pi}\) compatible with CMOS driving, while optimising both bandwidth and insertion loss. As reported in Table~1, the proposed method enables the identification of fabrication-tolerant designs that fulfil the specified performance metrics for both bottom and top electrode configurations. It is worth noting that, given the considered geometry and parameters, for comparable \(V_{\pi}\) values, bottom electrodes perform better in terms of insertion loss, while top electrodes show a slight advantage in terms of 3~dB EOE bandwidth. 
An intuitive physical explanation can be given by considering the dominant loss mechanisms. For bottom electrode configurations, when the SiN and oxide thickness are sufficiently large, metal-induced absorption becomes negligible, and insertion loss is mainly determined by tapering loss, which is independent of the electrode gap. This enables low \(V_{\pi}\) designs with limited variability and minimal propagation loss. In contrast, for top electrode configurations, the metal lies in the same plane as the optical mode, leading to non-negligible propagation losses even for larger electrode gaps. As a result, an additional insertion loss penalty is generally observed. While similarly low \(V_{\pi}\) values can be achieved, this typically requires larger electrode gaps as for bottom electrode, which in turn favours higher bandwidth. For the cases considered here, no significant difference in robustness to fabrication tolerances is observed between bottom and top electrode configurations. The corresponding geometric parameters of the modulators are reported in the Supplementary Material. 

Although, based on the cases considered, bottom-electrode architectures are preferable when minimising insertion loss is the primary objective, whereas top-electrode configurations offer advantages when targeting higher bandwidth, the choice of architecture should also take fabrication considerations into account. In particular, bottom-electrode designs can potentially rely on metal routing and bond pads fabricated in the back-end-of-line of a CMOS foundry, thereby reducing the heterogeneous integration of LN on silicon photonics to the MTP of LN waveguides. In contrast, even when metal electrodes are transferred with the LN waveguide, top-electrode implementations typically require additional post-processing steps to realise metal routing and bond pads. Furthermore, it is challenging to fabricate copper on top of LN meaning that in practice an alternative

\section{Discussion and Conclusion}

In this work, we presented a variability-aware design and optimisation framework for heterogeneously integrated lithium niobate traveling-wave modulators targeting wafer-scale fabrication. Through a comparative analysis of top- and bottom-electrode architectures, we identified the dominant geometric parameters governing electro-optic efficiency, optical propagation loss, RF performance, and
tapering loss, as well as their nonlinear interaction effects. The combination of global correlation analysis and Sobol variance decomposition revealed that, while certain performance metrics are primarily driven by single-parameter variations, others depend strongly on second-order interactions. Furthermore, some design parameters have opposing effect on different performance metrics,  inherent trade-offs. These findings emphasize the limitations of sequential or single-metric design approaches and motivate the
need for simultaneous multi-parameter co-optimisation.

Building on this analysis, we introduced a statistics based optimisation strategy that explicitly incorporates fabrication-induced variations into the design loop. By combining Monte Carlo sampling with a unified cost function that penalises both deviations from the target performance and sensitivity to geometric fluctuations, and by solving the resulting optimisation problem using the gradient-free CMA-ES algorithm, we demonstrate the ability to identify modulator geometries that, for a given device length, simultaneously achieve the desired \(V_{\pi}\), insertion loss, and bandwidth, while exhibiting improved robustness. Case studies illustrate that designs with similar nominal performance can exhibit substantially different variability behavior, highlighting the importance of robustness-aware optimisation in the context of wafer-scale manufacturing.

The proposed framework is grounded in experimentally measured fabrication statistics obtained from a MTP pilot line, including realistic
parameter distributions that are directly embedded in both the performance model and the cost function $J(\mathbf{x})$. Consequently, the optimisation results reflect the current state of the fabrication process rather than idealised or assumed tolerances. The optimized geometries obtained within this framework define concrete, pilot-line-compatible designs that can be fabricated and evaluated experimentally.

Future work will extend the present simulation database beyond 1310\,nm to other wavelengths, as well as to other material stacks and alternative metallisation schemes. Furthermore, advanced traveling-wave electrode geometries, including T-shaped electrodes, will be incorporated to enable longer modulator devices with reduced RF loss and bandwidths exceeding 110\,~GHz, while maintaining compatibility with wafer-scale heterogeneous integration flows. 

Rather than replacing experimental validation, the proposed framework is intended
to operate within an iterative design--fabrication--verification loop. The
sensitivity and interaction information extracted from the model enables to make
targeted, low-dimensional design-of-experiments focused on a limited set of
high-impact process parameters. Wafer-level measurement data can then be used to
verify model predictions, monitor process evolution, and update the statistical
distributions used in subsequent optimisation cycles. This iterative approach
supports progressive refinement of both device designs and process assumptions,
facilitating convergence toward stable and reproducible modulator performance
under pilot-line manufacturing conditions.

From a broader system-level perspective, no single modulator design is universally optimal, as different applications impose distinct priorities on device length, insertion loss, bandwidth, drive voltage, and robustness. By explicitly navigating these trade-offs under realistic fabrication constraints, the proposed framework enables the deliberate tailoring of modulator designs to specific application requirements and customer needs, rather than forcing convergence to a single, ideal design. As heterogeneous photonic integration moves toward high-volume manufacturing, such variability-aware design methodologies will be essential for bridging the gap between state-of-the-art single-device performance and reproducible, high-yield production.

\section{Supplementary Material}
The supplementary material contains three sections. The first section explains the methodology of data collection and provide some theoretical explanation on how the performance
metrics were calculated. The second section provides some theoretical background and describes computational methodology for the global sensitivity analysis presented
in Section 3. The third section explains the optimisation pathway in a bit more detail and shows the geometric parameters of the example designs used in Fig 4., Fig 5. and Table I.

\begin{acknowledgments}
We would like to thank Prof. Peter Bienstman for his valuable input. The authors would like to also thank the imec-Leuven teams providing the silicon and silicon nitride photonic waveguide circuits for our pilot line. The research has been made possible by FWO and F.R.S.-FNRS under the Excellence of Science
(EOS) program (40007560). The work has been supported by The Dutch National Growth Fund PhotonDelta, CHIPS-JU PhotonixFAB (101111896) and STARLIGHT(101194170). 
\end{acknowledgments}

\nocite{*}
\bibliography{Paper_references}

\end{document}